\documentclass[journal]{IEEEtran}
\usepackage{graphicx}
\usepackage{epsf}
\usepackage{mathtools}
\usepackage{amssymb}
\usepackage{amsmath, amsfonts}
\usepackage{latexsym}
\usepackage{multirow}
\usepackage{multicol}
\usepackage[table]{xcolor}
\usepackage{color}

\usepackage{tabularx}
\usepackage{lipsum}

\usepackage{mathrsfs}

\usepackage{fixltx2e}

\usepackage[mathscr]{euscript}

\usepackage{commath}
\usepackage{MnSymbol}

\usepackage{mathtools} 
\DeclarePairedDelimiterX\setc[2]{[}{]}{\,#1 \;\delimsize\vert\; #2\,}
\DeclarePairedDelimiterX\parth[2]{(}{)}{\,#1 \;\delimsize\vert\; #2\,}

\DeclarePairedDelimiter{\round}{\lfloor}{\rceil}

\PassOptionsToPackage{hyphens}{url}
\usepackage[unicode=true, bookmarks=true, bookmarksnumbered=true, bookmarksopen=true, bookmarksopenlevel=1, breaklinks=false, pdfborder={0 0 0}, pdfborderstyle={}, backref=false, colorlinks=false]{hyperref}

\usepackage[ruled,linesnumbered]{algorithm2e}

\usepackage{theorem}
{
\theorembodyfont{\rmfamily}

\newtheorem{remark}{Remark}

\newtheorem{proposition}{Proposition}
\newtheorem{definition}{Definition}

}

\usepackage{subcaption}
\usepackage{bbm}

\definecolor{orange}{RGB}{255,127,0}
\definecolor{blue}{RGB}{0,0,255}
\definecolor{red}{RGB}{255,0,0}
\definecolor{green}{RGB}{50,160,50}
\definecolor{grey}{RGB}{125,120,125}

\begin{document}
{
\title{{\fontsize{17}{4}\selectfont Reinforcement Learning for Accident Risk-Adaptive V2X Networking}}

\author
{
Seungmo Kim, \textit{Member}, \textit{IEEE}, and Byung-Jun Kim

\vspace{-0.3 in}

\thanks{S. Kim is with the Department of Electrical and Computer Engineering, Georgia Southern University in Statesboro, GA, USA. B. J. Kim is with Department of Statistics at Virginia Tech in Blackaburg, VA, USA. The corresponding author is S. Kim: contact at seungmokim@georgiasouthern.edu.}
}

\maketitle
\begin{abstract}
The significance of vehicle-to-everything (V2X) communications has been ever increased as connected and autonomous vehicles (CAVs) get more emergent in practice. The key challenge is the dynamicity: each vehicle needs to recognize the frequent changes of the surroundings and apply them to its networking behavior. This is the point where the need for machine learning is raised. However, the learning itself is extremely complicated due to the dynamicity as well, which necessitates that the learning framework itself must be resilient and flexible according to the environment. As such, this paper proposes a V2X networking framework integrating reinforcement learning (RL) into scheduling of multiple access. Specifically, the learning mechanism is formulated as a multi-armed bandit (MAB) problem, which enables a vehicle, without any assistance from external infrastructure, to (i) learn the environment, (ii) quantify the accident risk, and (iii) adapt its backoff counter according to the risk. The results of this paper show that the proposed learning protocol is able to (i) evaluate an accident risk close to optimal and, as a result, (ii) yield a higher chance of transmission for a dangerous vehicle.
\end{abstract}


\section{Introduction}\label{sec_intro}

\subsection{Background}
\subsubsection{Significance of V2X Communications}
Vehicle-to-Everything (V2X) communications have the potential to significantly bring down the number of vehicle crashes, thereby reducing the number of associated fatalities \cite{gdot_1}. The capability gave V2X communications the central role in constitution of intelligent transportation system (ITS) for connected and autonomous vehicles (CAVs).

\subsubsection{Challenge}
Due to the very high mobility of vehicles, there will be frequent change in the topology of the network. Mainly, routing, mobility and security are affected by this frequent topology change \cite{icc19}.

Analysis of such ever-dynamic environments surrounding a vehicle has been attempted in the literature of \textit{usage-based insurance (UBI)}. This technique is a means to incorporate analysis of the driving behaviors (e.g., speed, mileage, and harsh braking/accelerating) into determination of an auto insurance rate. The best practice that has been adopted by many insurance programs to protect users' location privacy is the use of driving speed rather than Global Positioning System (GPS) data \cite{allstate18}.

However, the current UBI-based approaches have limitations to be applied to make a V2X network adapt itself to a dynamic environment. In practice, many environmental factors, such as traffic amount/pattern and regulations, influence the risk level to which a vehicle is exposed in ``real-time.'' Nevertheless, the current UBI schemes cannot reflect dynamic traffic patterns and environmental change.

\subsubsection{Approach Adopted by This Paper}
In a CAV environment, the vehicles must be able to explore, learn from, make accurate evaluation of the level of an accident risk, and adapt their transmission of basic safety messages (BSMs) to the environment. To date, there is no proposal fostering such a capability at each vehicle. This makes a compelling case that a V2X network should employ a machine learning functionality to keep itself nimble on the fly without the need for external update, which causes additional delay.

\textit{Online machine learning} is a method in which data becomes available in a \textit{streaming} manner \cite{ms19}, as opposed to batch (or offline) learning which is trained by an entire training data set at once. As such, online learning is known to be particularly efficient in areas where it is computationally infeasible to train over the entire dataset, which includes V2X environments.

Notice that this paper focuses on defining the level of ``accident risk'' that will be used as a metric, to which a V2X networking protocol will be adapted. There are a plethora of factors determining an accident risk, which makes the training data ``streamed'' over time. Hence, being able to dynamically adapt to new inputs, an online learning algorithm will very well suit this paper's learning model.

Furthermore, our proposed learning framework must be resilient. We adopt \textit{reinforcement learning} so the learning framework can enhance itself in the dynamic V2X environments. In particular, in order to reflect the high dynamicity, this paper characterizes the learning framework as a \textit{contextual multi-armed bandit (MAB)} problem.

\subsection{Contributions of This Paper}
This paper presents the following contributions:
\begin{enumerate}
\item \textit{Extendibility of the proposed learning framework:} While this paper selects (i) variance of speed and (ii) inter-vehicle distance as contexts to be learned, the proposed learning framework is designed in such a way that it can be easily extended to (i) accommodate any other accident risk-related contexts as ``online'' inputs and (ii) adapt the networking behavior to an ever-changing environment.
\item \textit{Generality of the accident risk measurement model:} While the existing UBI models consider driver-oriented factors only, this paper proposes an accident risk measurement model that can be extended to any factors to be learned: not only driver-related factors but external factors such as road condition and weather as well. It makes the proposed model far more generally applicable than the typical UBI models.
\item \textit{Capability of completely autonomous learning and networking adaptation:} The proposed protocol can be applied to any ``distributed'' and ``listen-before-talk (LBT)'' type of V2X network--e.g., dedicated short-range communications (DSRC) and cellular V2X (C-V2X) mode 4. As such, it is a remarkable improvement from the traditional UBI in such a way that each vehicle is able to learn the environment according to the context and hence improve the UBI model itself on the fly without external update nor prior knowledge.
\end{enumerate}

\begin{table}[t]
\caption{Key abbreviations and notations}
\centering
\begin{tabular}{ |l|l|}
\hline
\textbf{\cellcolor{gray!30}Abbreviation} & \cellcolor{gray!30}\textbf{Description}\\
\hline\hline
BSM & Basic safety message\\
C-V2X & Cellular V2X\\
DSRC & Dedicated short-range communications\\
EXP & Packet expiration\\
LBT & Listen-before-talk\\
MAB & Multi-armed bandit\\
PPP & Poisson point process\\
RL & Reinforcement learning\\
UBI & Usage-based insurance\\
VANET & Vehicular ad-hoc network\\
V2X & Vehicle-to-everything communications\\
\hline
\hline
\textbf{\cellcolor{gray!30}Notation} & \cellcolor{gray!30}\textbf{Description}\\
\hline\hline
$w_{j}$ & Weight for context in context $j$, $\mathbf{c}_{j}$, in $\mathcal{AR}_{t,i}$\\
$\mathcal{AR}_{t,i}$ & Accident risk of vehicle $i$ at time instant $t$\\
$\mathsf{bo}$ & Backoff counter, i.e., $\left[0, \text{CW}-1\right]$\\
$\mathbf{c}_{i}$ & Vector of ``context'' $i$, i.e., $\left(c_{v,i}, c_{d,i}\right)$\\
CW & Contention window\\
$d^{\text{min}}_{i}$ & Minimum inter-vehicle distance at vehicle $i$\\
$\Psi$ & Variance of speed from the road's speed limit\\
$\mathsf{N}_{\text{arm}}$ & Number of arms\\
$\mathcal{S}_{\text{ctx}}$ & Set of contexts, i.e., $\mathcal{S}_{\text{ctx}} = \left\{ v, d \right\}$\\
$\mathsf{bo}_{t,i}$ & Backoff value for vehicle $i$ at time $t$\\
$v_{i}$ & Speed of vehicle $i$\\
$\mathbf{w}_{i}$ & An arm of the bandit $i$, i.e., weights $\left( w_{v}, w_{d} \right)$\\
\hline
\end{tabular}
\label{table_abbreviations}
\end{table}

\section{Related Work}\label{sec_related}

\subsubsection{Analysis of V2X Networking Performance}
It is with particular challenge to analyze the performance of a V2X network. Analysis frameworks based on stochastic geometry for DSRC have been proposed \cite{arxiv19}-\cite{arxiv20}. They commonly rely on the fact that uniform distributions of nodes on $X$ and $Y$ axes of a Cartesian-coordinate two-dimensional space yield a Poisson point process (PPP) on the number of nodes in the space \cite{haenggi05}. This paper also applies the stochastic geometry framework for analysis of the proposed mechanism.

\subsubsection{Accident Risk Quantification}
As has been mentioned earlier, the UBI is the main literature where the most effort on accident risk quantification has been found \cite{survey18}. The technique is quite stable and reliable, which makes itself widely applied in practice already by the commercial insurance providers \cite{allstate18}.

Nevertheless, as pointed out in Section \ref{sec_intro}, the key limitation of the current literature on UBI is that only driver-related factors are taken into consideration. There are too many other factors that should be considered in order to accurately model an accident risk. Even further, the type, number, and influence in concert of the factors are changing all over time. Yet the existing proposals are not flexible enough to precisely characterize such dynamicity.

\subsubsection{Machine Learning in V2X}
Several bodies of prior work have been found on application of machine learning to V2X networking: (i) detection of anomaly and misbehaviors for security of a V2X network \cite{icc19}\cite{elephant18}; (ii) beam selection \cite{tnet18} and situational awareness \cite{asilomar18} in millimeter wave bands; and (iii) transmission path optimization via obstruction detection by training using satellite images \cite{lett20}.

However, no contribution has been found on adapting the BSM prioritization and networking lightening. We highlight the significance of the network load lightening for the following two reasons: (i) there is increasing demand for V2X traffic as more connected and autonomous vehicles are emergent on the road; (ii) second, the spectrum contention for V2X communications are getting more severe especially in the 5.9 GHz band in the United States, due to the recent spectrum re-alignment by the Federal Communications Commission (FCC) \cite{nprm}.

\begin{figure}
\centering
\includegraphics[width = \linewidth]{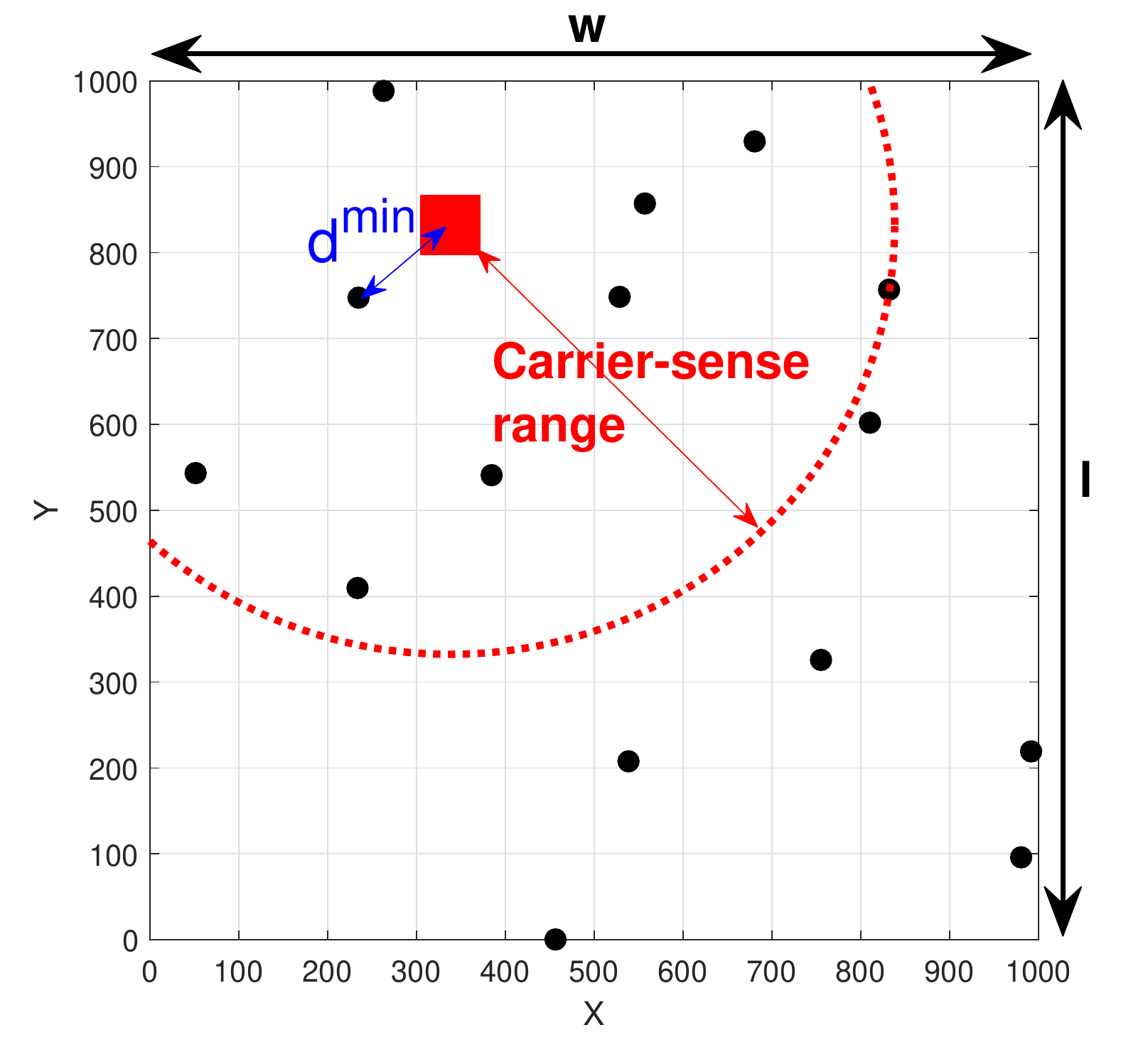}
\caption{Distribution of vehicles on $\mathbb{R}^2$ (with $l = w = $ 1e3 m and $\lambda = $ 15e-3 vehicles/m$^2$)}
\label{fig_scatter}
\end{figure}

\section{System Model}\label{sec_system_model}
\subsubsection{Network}
In the vehicular network model presented in this paper, a vehicle is assumed to be mobile. The network is completely \textit{distributed}, \textit{i.e.}, vehicular ad-hoc network (VANET), in which no central coordinator node nor infrastructure exists. Specifically, a \textit{distributed} type of V2X networks is assumed such as (i) DSRC \cite{11pstd} and (ii) mode 4 of cellular V2X (C-V2X) \cite{lbt19}, both of which are based on listen-before-talk (LBT). Also, exchange of information among vehicles is achieved via \textit{periodic broadcast} of BSMs.

Also, a network is \textit{fully connected}: every node is supposed to be equipped with communication functionality and hence is able to broadcast it whenever needed.

This system model can generally suit currently operating VANET systems in practice including IEEE 802.11-based system such as DSRC \cite{11pstd} and 802.11bd \cite{80211bd}. The model can also be applied to C-V2X mode 4 where the nodes are connected directly in a distributed manner without going through the network core, e.g., sidelink-based broadcast or groupcast as defined in the 3rd Generation Partnership Project (3GPP) Release 16 \cite{tr22886}. Commonly in the standards, a basic safety message (BSM) is \textit{broadcast without any feedback available}. This paper assumes 10 BSMs being broadcast per second.

\subsubsection{Geometry}
As shown in Fig. \ref{fig_scatter}, a two-dimensional space $\mathbb{R}^2$ is defined with the length and width of $l$ and $w$ meters (m), respectively. In order to capture a more dynamic and realistic movement of nodes in a vehicular network, this system model considers \textit{no separation of lanes}. Notice that such a generalized model enables the subsequent analyses more widely applicable \cite{access19}. Furthermore, to consider the most generic vehicle movement characteristic, this model assumes that every vehicle can move in any direction, which enables the system to capture every possible movement scenario including flight of UAVs, lane changing, intersection, and pedestrian walking.

The distribution of the nodes follows Poisson point process (PPP). This paper considers a homogeneous PPP $\Phi$ with the vehicle density $\lambda$. As such, the position of each node is uniformly distributed on each axis of X and Y between $\left[0, l\right]$ and $\left[0, w\right]$, respectively \cite{uniform}.

\section{Proposed RL Framework}\label{sec_proposed}
This section articulates the proposed RL algorithm enabling a vehicle to autonomously update its accident risk according to the environment.

As Fig. \ref{fig_overview} illustrates, in each time slot, a bandit (i.e., a vehicle) makes a selection among arms accounting the ``context'' that has been observed in the time slot. Each arm selection leads to quantification of (i) an accident risk level, (ii) backoff counter, (iii) successful transmission or EXP, and (iv) a reward. The vehicle learns from the rewards collected in the arms to make the choice of an arm in the current iteration. Over time, the vehicle's aim is to collect enough information about how the contexts and arms are mapped to each other, so that it can predict the next best arm to play by looking at the incoming context.

\subsection{Problem Formulation}
\subsubsection{MAB Problem}
In essence, the proposed RL framework is defined to learn $\mathbf{w}_{i} = \left( w_{v}, w_{d} \right)$ at each vehicle $i$ within a time slot $t$, in accordance with a context $\mathbf{c}_{i}$.

If vehicle $i$ (the bandit) knew the expected weight $\mathbf{w}_{i}$ for each context $\mathbf{c}_{i,t}$ in time slot $t$ a priori, it could simply (i) evaluate the accident risk for the given context based on (\ref{eq_AR}) and (ii) adjust its transmission priority in the network. However, since the vehicle $i$ does not know the environment, it has to learn the weight $\mathbf{w}_{i}$ over time.

In order to learn these values, vehicle $i$ has to try out different values for weight $\mathbf{w}_{i}$ on different contexts over time, which forms a \textit{MAB problem}. At the same time, it should ensure that those $\mathbf{w}_{i}$'s already proven to be accurate are used sufficiently often. Hence, the vehicle has to find a tradeoff between ``exploring'' $\mathbf{w}_{i}$'s, of which it has little knowledge, and ``exploiting'' the proven $\mathbf{w}_{i}$'s for high average learning performance. As such, the selection of the learning algorithm depends on the history of $\mathbf{w}_{i}$'s that have been selected in previous periods and the corresponding observed rewards, $\mathbf{r}_{i}$'s.

Let $\mathbf{c}_{i}$ denote a vector of contexts, which is formulated as
\begin{align}\label{eq_c}
\mathbf{c}_{i} = ( c_{v}, c_{d} )
\end{align}
where contexts $c_{v}$ and $c_{d}$ represent speed of each vehicle and inter-vehicle distance at each vehicle, respectively.

\subsubsection{Accident Risk}
We remind that a V2X environment is extremely dynamic presenting a wide variety of factors that cause an accident. By ``dynamic,'' we mean not only the \textit{quantity} of a factor but the \textit{type} of factors as well. As such, each vehicle needs to keep teaching itself how to deal with the factors ``in concert'' in such a dynamic environment. More specifically, in accordance with the environment, the relative significance of each factor should be differentiated. For instance, while a highly speeding vehicle marks a higher crash risk in a highway, a vehicle with a smaller inter-vehicle distance, $d^{\text{min}}_{i}$, poses a higher risk in a denser traffic.

While a plethora of factors must be considered, the proposed RL framework highlights its contribution by remaining flexible for any additional factor. Notice from (\ref{eq_c}) that as more contexts are considered to quantify an accident risk, the context vector $\mathbf{x}_{i}$ gets longer accordingly.

\begin{figure}
\centering
\includegraphics[width = \linewidth]{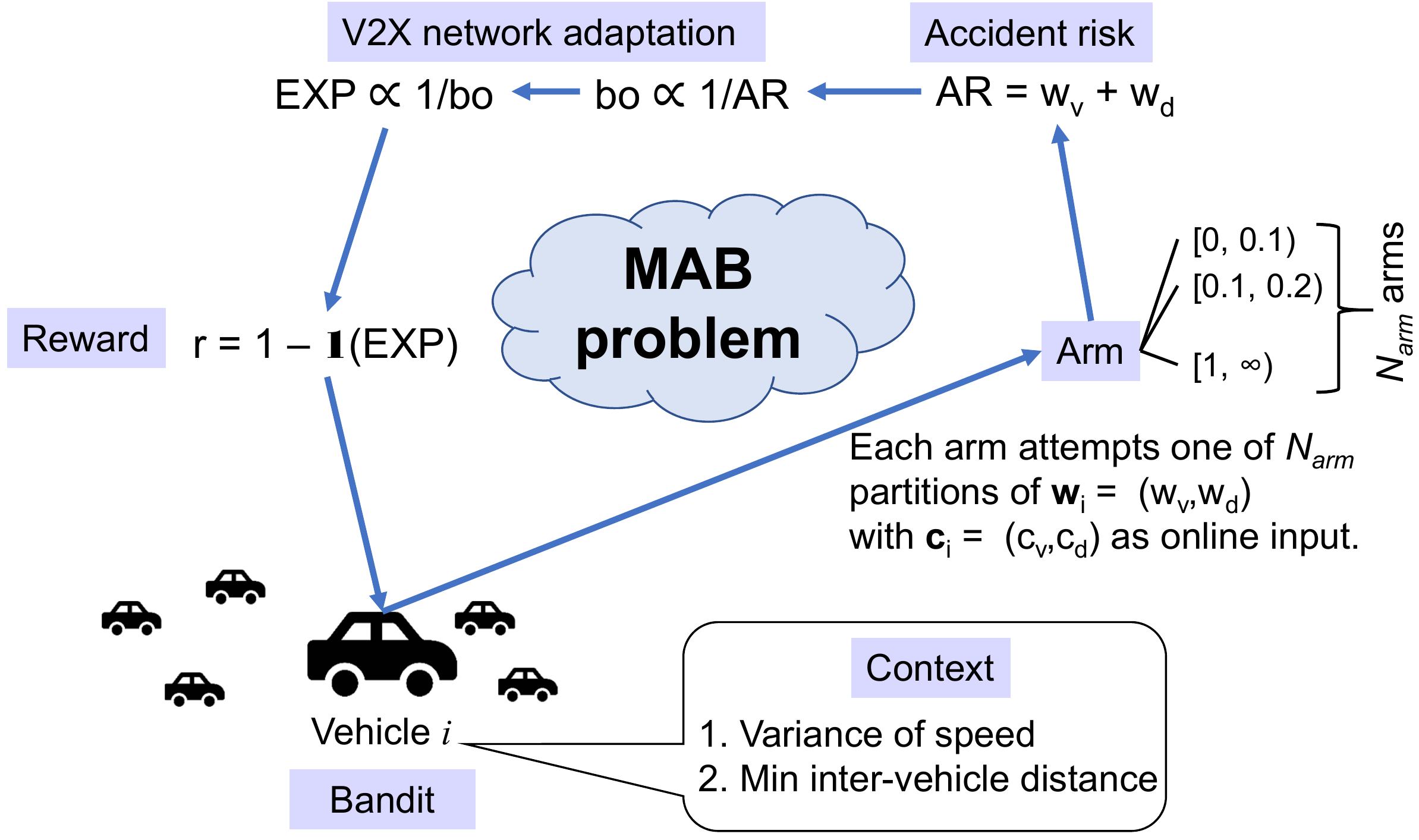}
\caption{A learning cyle of the proposed RL framework within a time slot $t$}
\label{fig_overview}
\end{figure}

\begin{definition}\label{definition_objective}
(Learning objective: Accident risk). \textit{Accident risk of vehicle $i$ at time instant $t$ is formulated as}
\begin{align}\label{eq_AR}
0 \le \mathcal{AR}_{t,i} = \displaystyle \sum_{j \in \mathcal{S}_{\text{ctx}}} w_{j,i} \le \mathsf{N}_{\text{arm}}
\end{align}
\textit{where $\mathcal{S}_{\text{ctx}}$ denotes the set of contexts (i.e., $\mathcal{S}_{\text{ctx}} = \left\{ v,d \right\}$ as shall be detailed in Definitions \ref{definition_cv} and \ref{definition_cd}). Since the coefficient $w$ for each arm ranges in $\left[ 0,1 \right]$, the quantity of an entire $\mathcal{AR}$ ranges in $\left[0, \mathsf{N}_{\text{arm}}\right]$.}
\end{definition}

\subsubsection{Contexts and Arm-Mapping Policies}
Now, one wishes to formulate the weights $w_{j,i}$ defined in (\ref{eq_AR}), which indicate the level of an accident risk at a vehicle.

The first type of context that this paper adopts is the \textit{variance of speed}. Operating speed at a vehicle has been found to be normally distributed with the mean being the speed limit of the road \cite{nhtsa09}. We take advantage of this premise to proceed to define our first accident risk factor.

The rationale behind this is as follows. First, it becomes dangerous being far from the mean. In other words, not only a too high speed, but is a too low speed dangerous. By defining as ``variance,'' a single quantity can express both. Second, a speed can be obtained at a vehicle without any external information. For instance, a speed limit is an easy number to obtain in many commercial GPS applications (e.g., Google Maps). Reliance on such an easily available quantity increases the applicability of the proposed algorithm.

\begin{definition}\label{definition_cv}
(Context 1: Variance of speed).
\begin{align}\label{eq_cv}
c_{v,i} = \left| v_{i} - v_{\text{ref}}\right|
\end{align}
where $v_{\text{ref}}$ gives the reference speed, i.e., the speed limit of the road in which vehicle $i$ is currently being driven.
\end{definition}

\begin{definition}\label{definition_wv}
(Optimal strategy for context 1: Mapping context $c_{v}$ to arm $w_{v}$). \textit{We remind that context $c_{v,i}$ is designed to be dangerous as the variance $c_{v,i}$ goes further from $v_{\text{ref}}$. As such, this paper proposes that the weight for this context, $w_{v,i}$, as}
\begin{align}\label{eq_wv}
w_{v,i} = \tanh \left( k_{v} c_{v,i} \right), \hspace{0.1 in} c_{v,i} \ge 0
\end{align}
\textit{where $k_{v}$ is a constant that controlls the steepness of $w_{v,i}$.}

\textit{It is interpreted as follows. The weight $w_{v,i}$ becomes neutral (i.e., 0.5) when vehicle $i$ keeps its speed $v_{i} = v_{\text{ref}}$, i.e., $c_{v,i} = 0$. As $w_{v,i}$ gets smaller/greater than 0.5, as the weight $w_{v,i}$ grows.}
\end{definition}

In addition to the speed, this paper adopts the \textit{inter-vehicle distance} as another critical factor in determining an accident risk at a vehicle \cite{dsafe}.

\begin{definition}\label{definition_cd}
(Context 2: Minimum inter-vehicle distance in reference to a safe braking distance).
\begin{align}\label{eq_cd}
c_{d,i} = \frac{d^{\text{min}}_{i}}{d_{\text{ref},i}}
\end{align}
\textit{where $d_{\text{ref},i}$ denotes the safe stopping distance, which is given by \cite{dsafe}}
\begin{align}
d_{\text{ref},i} = r v_{i} \frac{10}{36} + \frac{v_{i}^2}{b}
\end{align}
\textit{where $v_{i}$ is measured in km/h; $r$ gives the driver reaction time (secs), which is set to 1.5 secs in this paper; and $b$ indicates the braking coefficient factor, which is nominally set to 170 assuming dry, level pavement.}

\textit{Also, the first term in (\ref{eq_cd}) is defined as}
\begin{align}
d^{\text{min}}_{i} = \min_{i \neq j, \hspace{0.02 in} j \in \mathcal{S}_{\text{nbr}} } d_{i,j}
\end{align}
\textit{where $\mathcal{S}_{{\rm{nbr}}}$ is the set of neighboring vehicles.}
\end{definition}

\begin{figure}
\centering
\includegraphics[width = \linewidth]{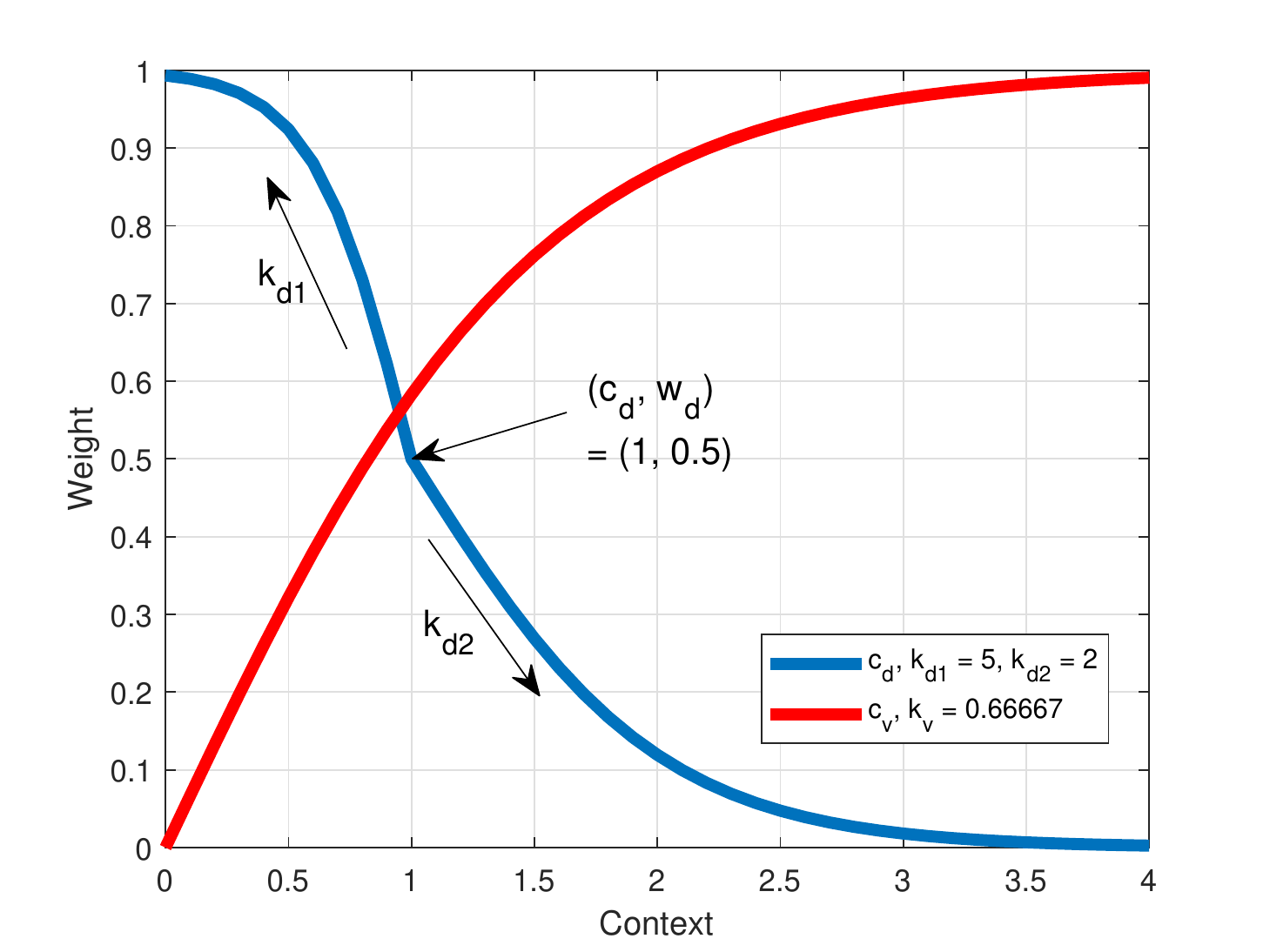}
\caption{Weights as functions of contexts $c_{v,i}$ and $c_{d,i}$ (with coefficients $k_{v}$ = 2/3 and $k_{d}$ = 5)}
\label{fig_w_vs_c}
\end{figure}

\begin{definition}\label{definition_wd}
(Optimal strategy for context 2: Mapping context $c_{d}$ to arm $w_{d}$). \textit{Recall that context $c_{d,i}$ is designed to be (i) dangerous when $c_{d,i} < 1$, i.e., $d^{\text{min}}_{i} < d_{\text{ref}}$; (ii) safe while being kept $c_{d,i} > 1$, i.e., $d^{\text{min}}_{i} > d_{\text{ref}}$; and (iii) on the border if equal to 1, i.e., $d^{\text{min}}_{i} = d_{\text{ref}}$. In accordance, this paper proposes to allocate the weight for this context, $w_{d,i}$, as a sigmoid, which is formally written as}
\begin{align}\label{eq_wd}
w_{d,i} = \frac{1}{1 + e^{k_{d}\left(c_{d,i}-1\right)}}, \hspace{0.1 in} c_{d,i} \ge 0
\end{align}
\textit{where $k_{d}$ is a constant that controlls the steepness of the weight function.}

\textit{It is interpreted as follows. The weight $w_{d}$ becomes neutral (i.e., 0.5) when vehicle $i$ keeps $d^{\text{min}}_{i} = d_{\text{ref}}$. Now, $w_{d}$ gets (i) smaller than 0.5 as $c_{d}$ goes negative and (ii) greater than 0.5 as $c_{d}$ goes positive.}
\end{definition}

\begin{remark}\label{remark_w_vs_c}
(Weight versus context). \textit{Fig. \ref{fig_w_vs_c} demonstrates the allocation of $w$ versus $c$ as have been proposed in Definitions \ref{definition_wv} and \ref{definition_wd}. As a remark, we remind that (i) $c_{v}$ is considered the safest when the current speed is equivalent to the speed limit of the current road, which allocates $w_{v} = 0$ indicating ``no accident danger''; and (ii) $c_{d}$ becomes safer as the value of $c_{d}$ gets greater.}

\textit{In particular, for context $c_{d}$, $w_{d}$ becomes 0.5 at $c_{d} = 1$, the point at which the context $c_{d}$ becomes not either safe nor dangerous. Also, an inflection point on $w_{d}$ is noteworthy: it shows that one can control the steepness of a weight function by varying the value of the coefficient $k_{d}$.}
\end{remark}

\begin{algorithm}[t]
\SetAlgoLined
Input: $T$, $\textbf{c}_{i}$, $\epsilon$

\For{t = 1, $\cdots$, T}{
\eIf{$t \le T_{\text{trn}}$}{
\textbf{``Training''}\\
Observe context $\mathbf{c}_{i}$\\
Give a random value for an arm $\mathbf{w}_{i}$\\
Compute the danger $\mathcal{AR}$\\
Allocate a backoff value according to $\mathcal{AR}$\\
Compute reward $\mathbf{r}_{i}$
}{
\eIf{rand $\le \epsilon$}{
\textbf{``Explore''}\\
Observe context $\mathbf{c}_{i}$\\
Repeat Lines 5 through 9
}{
\textbf{``Exploit''}\\
Observe context $\mathbf{c}_{i}$\\
Select an arm $\mathbf{w}_{i}$ based on (\ref{eq_wi_xt})\\
Repeat Lines 7 through 9
}
}
}
\caption{Proposed Algorithm for Learning and Network Adaptation}
\end{algorithm}

\subsection{Algorithm for Learning and Network Adaptation}
\subsubsection{Overview of the Proposed Algorithm}
In each discrete time slot $t$, the bandit (i.e., arbitrary vehicle $i$ in the network) performs the proposed algorithm to (i) learn the context in the changing environment and (ii) be given an opportunity of transmitting a BSM in accordance with the result of the learning. A detailed procedure follows as below.

The bandit is given a certain length of \textit{training phase} at the beginning. In a training phase, upon observation of a certain context, an arm is chosen uniformly randomly among the $\mathsf{N}_{\text{arm}}$ arms. The weight for the context selected via the selected arm is used to compute a value for $\mathcal{AR}$.

After the training is completed, the bandit still goes into an \textit{explore} operation with a certain probability $\epsilon$. When exploring, the same operation is performed as in a training phase. However, when the bandit \textit{exploits} with the probability of $1 - \epsilon$, it selects an arm $\mathbf{w}_{t,i}$, referring to the history of $\mathbf{r}$ that has been collected so far, as shall be formulated in (\ref{eq_wi_xt}).

\subsubsection{Explore vs. Exploit}
We formulate the \textit{exploit} and \textit{explore} as follows.
\begin{proposition}\label{proposition_exploit}
($\mathbf{w}_{i}$ when exploited). \textit{When vehicle $i$ exploits at a time instant $t = t_{0}$, the bandit selects the arm satisfying:}
\begin{align}\label{eq_wi_xt}
\mathbf{w}_{\text{xt},i} = \max_{t < t_{0}} \big( \mathbf{w}_{t,i} | \text{c}_{i} \big)
\end{align}
\end{proposition}

One can compare (\ref{eq_wi_xt}) to the bandit's \textit{exploring} among the arms, which can be formally written as
\begin{align}\label{eq_wi_xe}
\mathbf{w}_{\text{xe},i} = \mathcal{U}\left( 1, \mathsf{N}_{\text{arm}}\right)
\end{align}
where $\mathcal{U}\left( a,b \right)$ denotes a discrete uniform distribution with the interval of $\left[a, b\right]$ where $b > a$.

\begin{figure*}
\centering
\begin{subfigure}{.45\textwidth}
\centering
\includegraphics[width = \linewidth]{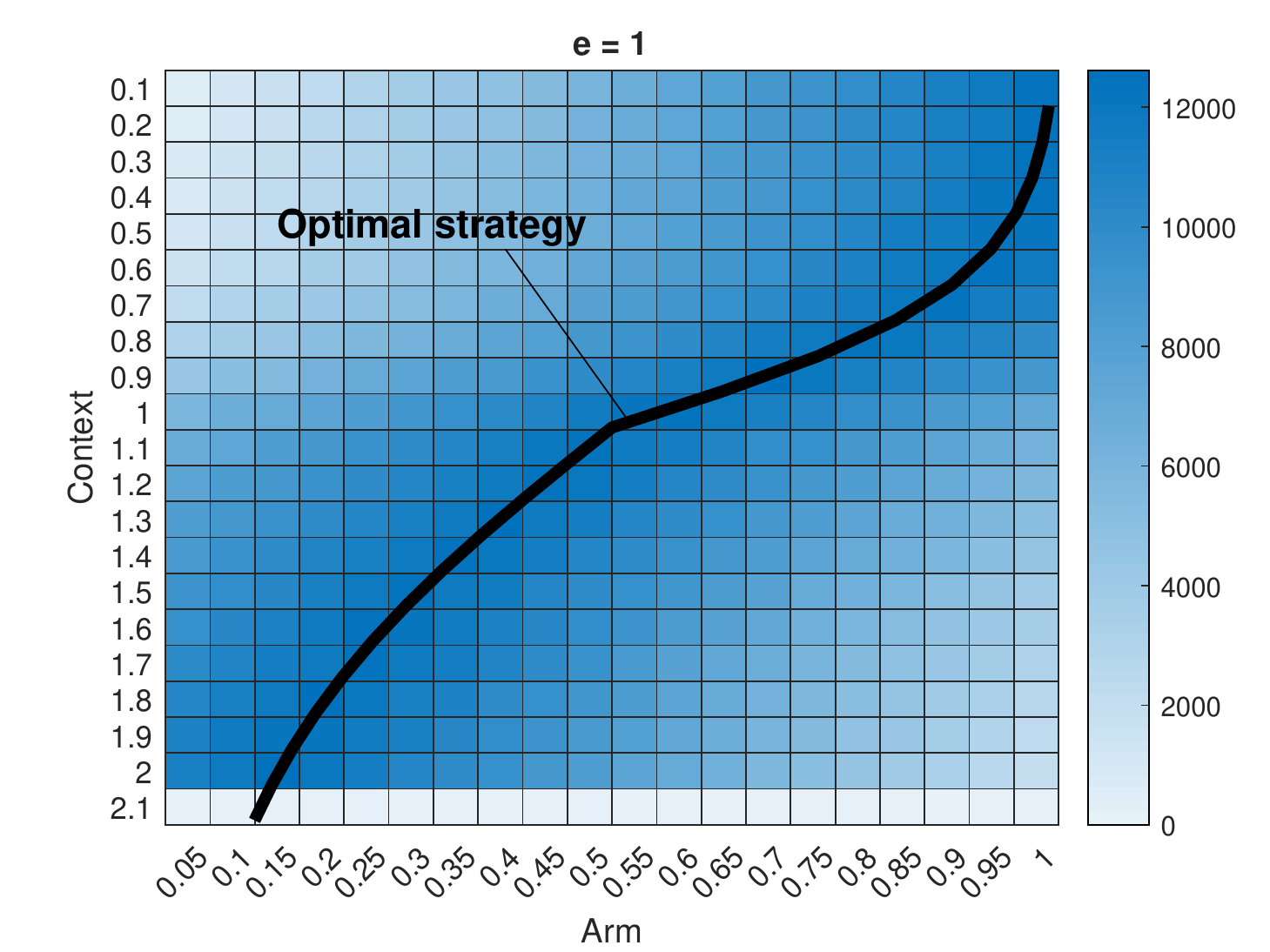}
\caption{A/B testing (i.e., $\epsilon = 1$)}
\label{fig_heatmap_e1}
\end{subfigure}\hspace{0.5 in}
\begin{subfigure}{.45\textwidth}
\centering
\includegraphics[width = \linewidth]{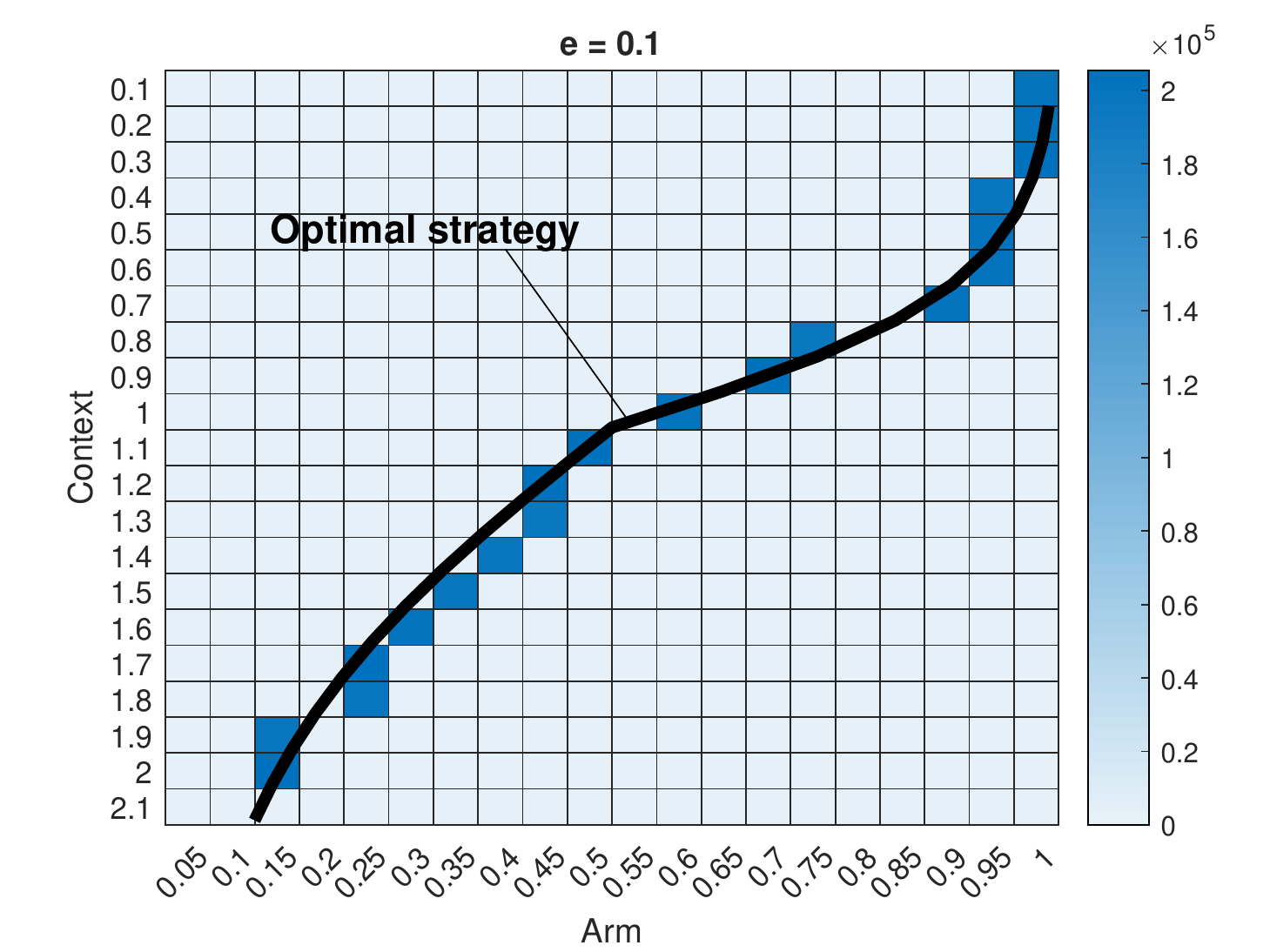}
\caption{$\epsilon$-greedy (with $\epsilon = 0.1$)}
\label{fig_heatmap_e01}
\end{subfigure}
\caption{Distribution of reward ($r_{d,i}$) versus arm ($w_{d,i}$) and context ($c_{d,i}$) according to MAB strategy (with $T = 10^{5}$ and $T_{\text{trn}} = T/10$)}
\label{fig_heatmap}
\end{figure*}

\subsubsection{Network Adaptation Based on the Learning Result}
We remind that the proposed protocol intends a more dangerous vehicle (i.e., with a higher $\mathcal{AR}$) to take a higher priority in a multiple access process. Thus, a backoff value in a CSMA process is allocated to be inversely proportional to the value of $\mathcal{AR}$, which is given by

The proposed protocol uses the packet expiration (EXP) rate as the metric associated with the reward in MAB. The EXP rate can be obtained within each vehicle completely autonomously. Specifically, according to the result of the learning, the vehicle (i.e., the bandit) calculates a backoff counter and allocates the counter to itself.

\begin{definition}\label{definition_bo}
(Backoff counter allocation according to $\mathcal{AR}$).
\begin{align}\label{eq_bo}
\mathsf{bo}_{t,i} = \round*{ \frac{1 - \text{CW}}{\mathbb{N}\left[ \mathcal{S}_{\text{ctx}} \right]} \mathcal{AR}_{t,i} + \text{CW} - 1 }
\end{align}
\textit{where $\round{\cdot}$ denotes rounding to the closest integer.}

\textit{Note that function $\mathsf{bo}_{t,i}$ is designed to become (i) the maximum (i.e., $\text{CW} - 1$) at the minimum $\mathcal{AR}_{t,i}$ (i.e., 0) and (ii) the minimum (i.e., $0$) at the maximum $\mathcal{AR}_{t,i}$ (i.e., $\mathsf{N}_{\text{arm}}$).}

\textit{With (\ref{eq_bo}), one can easily distinguish the proposed backoff allocation scheme with the traditional one where a backoff value is allocated completely uniformly.}
\end{definition}

As has been illustrated in Fig. \ref{fig_overview}, a reward is granted according to the result of a packet transmission at vehicle $i$, which is formally written as
\begin{definition}\label{definition_r}
(Reward according to packet transmission success or failure).
\begin{align}\label{eq_r}
\mathbf{r}_{t,i} = 1 - \mathbbm{1}\left(\text{EXP}\right),
\end{align}
\textit{which means that if vehicle $i$ succeeds to transmit a packet (i.e., no EXP), a reward of 1 is granted. We assume that a success in transmission does not guarantee a successful reception, which is a key characteristic in a distributed V2X network such as DSRC and C-V2X mode 4.}
\end{definition}

\subsubsection{Performance Measurement Metrics}\label{sec_proposed_metrics}
We proceed to evaluate the performance of the proposed mechanism. There are two separate metrics for quantifying the performance of (i) learning and (ii) networking.

The learning performance is measured via the \textit{regret of learning} \cite{ms19}. Specifically, it is defined as the expected difference in the amount of received data achieved (i) via an optimal way (i.e., a priori knowledge on the optimal $\left(w_{v}, w_{d}\right)$) and (ii) via the proposed RL algorithm. The regret after $T$ rounds of learning at vehicle $i$ is formally written as
\begin{align}\label{eq_rho}
\rho_{i}\left(T\right) &= \mathbb{E}\left[ \displaystyle \sum_{t=1}^{T} \bigg( \mathbf{r}^{\ast}\left(\mathbf{c}_{t,i}\right) - \mathbf{r}\left(\mathbf{c}_{t,i}\right) \bigg) \right]\nonumber\\
&= \displaystyle \sum_{t=1}^{T} \mathbb{E}\left[ \mathbf{r}^{\ast}\left(\mathbf{c}_{t,i}\right) \right] - \sum_{t=1}^{T} \mathbb{E}\left[ \mathbf{r}\left(\mathbf{c}_{t,i}\right) \right]
\end{align}
where $\mathbf{r} = \left(r_{v}, r_{d}\right)$ and $\mathbf{r}^{\ast}$ denotes the reward associated with an optimal strategy.

For evaluating the networking performance, we define the \textit{probability that a packet is transmitted}, which is denoted by $\tau$ \cite{arxiv19}. Specifically, $\tau$ is the probability that a packet goes through a backoff process, which is (i) initiated by allocation of a backoff counter and (ii) processed longer or shoter depending on the level of competition for the medium within the vehicle's carrier-sense range.

\section{Numerical Results}\label{sec_numerical}

\subsection{Setting}

\begin{table}[hbtp]
\caption{Key parameters}
\centering
\begin{tabular}{ |l|l|}
\hline
\textbf{\cellcolor{gray!30} Parameter (symbol)} & \cellcolor{gray!30}\textbf{Value}\\
\hline\hline
\# arms ($\mathsf{N}_{\text{arm}}$) & 100\\
\# periods ($T$) & 10$^5$ secs\\
Length of training phase ($T_{\text{trn}}$) & $T/10$ secs\\
Reference speed ($v_{\text{ref}}$) & 60 mph\\
CW & 15\\
\hline
\end{tabular}
\label{table_parameters}
\end{table}

This section demonstrates the results of the proposed framework. Specifically, numerical computations were performed to evaluate the proposed (i) RL and (ii) networking performances. The performances were measured at an arbitrary vehicle $i$ in a network. Also, notice that we employ a numerical solve for evaluation of $\tau$, which has been mentioned in Section \ref{sec_proposed_metrics}, due to massive computation load caused by recursiveness in a Markov process \cite{arxiv19}. Key parameters and the values are summarized in Table \ref{table_parameters}.

This section shows the results on context $c_{d,i}$ only, but it is obvious that the same numerical approaches can be applied to the other context $c_{v,i}$, because the proposed RL framework learns the significance $w_{d,i}$ and $w_{v,i}$ independently. Notice that this RL framework design enables even further extension to accommodation of other contexts for learning their weights to more accurately measure the accident risk $\mathcal{AR}$.

\begin{figure}
\centering
\includegraphics[width = \linewidth]{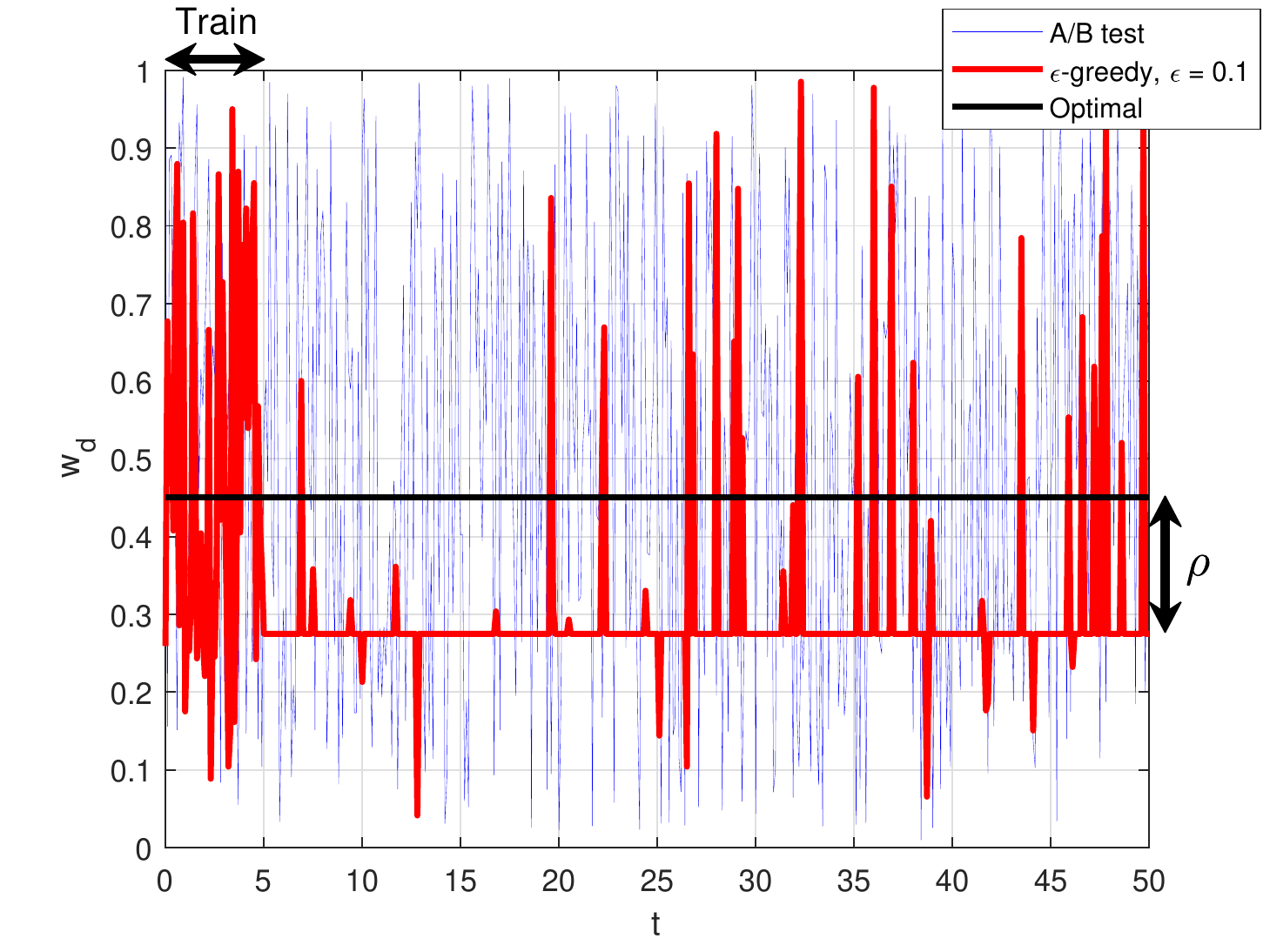}
\caption{Time convergence of the proposed RL algorithm according to MAB strategy (with $c_{d} = 1.1$)}
\label{fig_convergence}
\end{figure}

\subsection{Results and Discussion}

\subsubsection{Performance of RL according to MAB Strategy}
Fig. \ref{fig_heatmap} provides heatmaps to demonstrate how the reward is distributed versus the context and arm. Figs. \ref{fig_heatmap_e1} and \ref{fig_heatmap_e01} compare different representative MAB strategies--i.e., A/B testing and $\epsilon$-greedy. One can observe from the comparison that a higher probability of ``exploitation'' (i.e., $1 - \epsilon = $ 0 and 0.9 in Figs. \ref{fig_heatmap_e1} and \ref{fig_heatmap_e01}, respectively) yields a clearer tendency that has been observed in Fig. \ref{fig_w_vs_c}.

\subsubsection{Performance of RL in terms of convergence}
Fig. \ref{fig_convergence} sets a certain value of context (i.e., $c_{d} = 1.1$) to focus on the time convergence. As such, the figure presents the convergence of the proposed algorithm according to the same types of MAB strategy as in Fig. \ref{fig_heatmap}. Notice that this plot is a result with a significantly short duration (i.e., $T = 50$ sec) just for observation of the convergence, which, in turn, is with $T_{\text{trn}} = T/10 = 5$ sec. (We remind that this is just an example: in practice, we suggest one to elongate the initial training time as more accident causing contexts are considered in (\ref{eq_AR}).) While an A/B test (blue line) remains to be random all through $T$, an $\epsilon$-greedy strategy (red line) explores only at a given probability (i.e., $\epsilon$ = 0.1). The optimal $w_{d}$ for $c_{d} = 1.1$ can be found from either Fig. \ref{fig_w_vs_c} or \ref{fig_heatmap}.

\begin{figure}
\centering
\includegraphics[width = \linewidth]{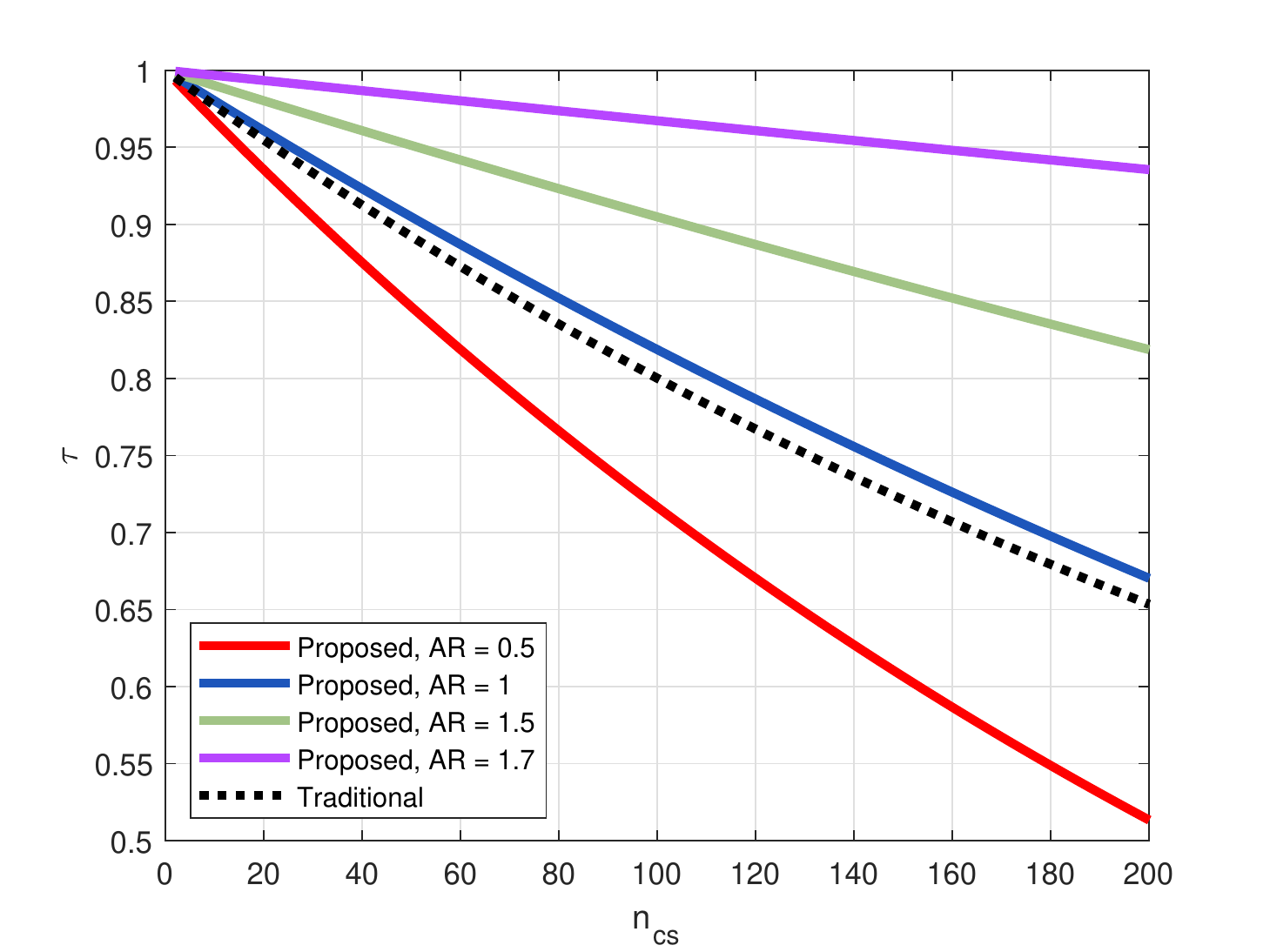}
\caption{Probability of a packet transmission at an arbitrary vehicle versus the number of competing vehicles (with CW $= 15$)}
\label{fig_tau}
\end{figure}

Also notice that the $\epsilon$-greedy strategy is not able to find a completely accurate arm as the optimal strategy does, mainly due to (i) a short training phase and (ii) discretization in the categorization of $c_{d}$ into a $w_{d}$. The regret $\rho$ is approximated to be 0.15 in the example shown in Fig. \ref{fig_convergence}. (See (\ref{eq_rho}) for formulation of $\rho$.)

\subsubsection{Performance of Networking}
Fig. \ref{fig_tau} compares $\tau$, the probability that a packet goes through the backoff process and is able to be transmitted, versus $n_{cs}$, the number of vehicles in vehicle $i$'s carrier-sense range. This result shows that the proposed protocol presented in Definition \ref{definition_bo} prioritizes a packet with a higher danger (i.e., a larger value for $\mathcal{AR}$), while a packet with a less accident risk marks a lower transmission probability. 

Notice the inverse relationship of EXP with $\tau$, which is given by $\mathbb{P}\left[\text{EXP}\right] = 1 - \tau$. As such, referring to (\ref{eq_r}), one can infer that the reward is piled faster with a larger $\tau$: i.e., $\mathbf{r}_{t,i} \propto \tau$. This relationship forms the chain between the result of a packet transmission linked to the vehicle's learning in the next round.

\section{Conclusions}
This paper proposed (i) a RL framework enabling a vehicle to autonomously adapt itself to dynamic environments and (ii) a V2X networking protocol prioritizing a vehicle with a high accident risk. This paper formulated the RL framework as a contextual MAB problem where the multiple arms are used to test all possible values for the ``weight'' of a certain context. The weight selected in a best arm was used for calculation of the accident risk at a vehicle in a time slot. The vehicle's priority of transmitting the current packet in a network was determined based on the accident risk. Whether the packet could be transmitted or not determined the reward, which was used in the selection of the next arm.

We foresee that this paper's finding will have direct impacts to the practice. For instance, the current autonomous driving functionalities (i.e., Auto Pilot by Tesla) can benefit from this framework. Our results showed that a vehicle could autonomously train itself on a policy of speed and inter-vehicle distance. This trained data could adapt the transmission priority according to the result of learning, which consequently increased the performance of a vehicle at a high crash risk.

Possible future extensions include performance evaluation of the proposed RL framework with more accident causing factors.


\end{document}